\documentclass[%
 reprint,
superscriptaddress,
 amsmath,amssymb,
 aps,
pra,
]{revtex4-2}

\usepackage{graphicx}
\usepackage{dcolumn}
\usepackage{bm}
\usepackage{bbm}
\usepackage{subcaption}
\usepackage{comment}
\usepackage{physics}
\usepackage{amsmath}
\usepackage{bbold}
\usepackage{caption}
\usepackage{xcolor}

\begin{document}

\preprint{APS/123-QED}

\title{Programmable integrated source of polarization and frequency-bin \\ hyperentangled photon pairs}
\author{Colin Vendromin}
\email{colin.vendromin@utoronto.ca}
\affiliation{Department of Physics, University of Toronto, 60 St. George Street, Toronto, Ontario, Canada M5S 1A7}%
\author{J. E. Sipe}
\affiliation{Department of Physics, University of Toronto, 60 St. George Street, Toronto, Ontario, Canada M5S 1A7}
\author{Marco Liscidini}
\affiliation{Department of Physics, University of Pavia, Via Bassi 6, 27100, Pavia, Italy}

\date{\today}
\begin{abstract}
We present a system capable of generating programmable polarization and frequency-bin hyperentangled photon pairs in an integrated photonic device. The structure is composed of ring resonators, each generating photon pairs with the same polarization in two pairs of frequency bins via spontaneous four-wave mixing. By combining several rings and controlling the amplitude and phase of the pump field, one can construct ``piece-by-piece" several hyperentangled states. For example, we consider a system composed of four rings and show that the density operator of the generated state describes a polarization and frequency-bin hyperentangled state. Finally, we calculate the expected generation rate and discuss the device efficiency.
\end{abstract}

\maketitle

\section{Introduction}
Photonic integrated circuits (PICs) enable the integration of multiple optical components, reducing the overall size and cost of optical systems, and leading to scalability beyond the limits of bulk optics \cite{ElshaariPICs2020, WangPICs2020, WangPICs2018, ElshaariPICs2017}. They offer precise control of light propagation, and through 
the design of waveguides and dielectric structures that enhance the light-matter interaction they can lead to an increased 
efficiency of 
non-classical light generation\cite{MahmudluOnChipQudit2023}. 

An important example
of non-classical light is a photon pair.
The entanglement of a pair of photons can be a 
key resource for quantum information processing, such as quantum computing \cite{TaballioneQC2023,QiangQC2018}, quantum cryptography \cite{Pseiner_2021,YinQKD2020}, and quantum teleportation \cite{LlewellynQtele2020}. Central to this resource is the degree of freedom (DOF) that is used to encode the information contained in the state. Polarization and time are usually the preferred DOFs for bulk optics. But
for PICs, arbitrary polarization states are difficult to 
manipulate, and the time DOF typically requires delay lines that are challenging to implement on-chip. Instead, the ability to control light propagation
in PICs makes the path DOF a natural choice \cite{QiangQC2018, CrespiPATH2013}. 

Alternatively, energy can be used as a DOF in the form of frequency bins, where the frequency entangled photons are routinely generated by a nonlinear process such as spontaneous parametric downconversion (SPDC) or spontaneous four-wave mixing (SFWM) in ring resonators \cite{AndreaSabattoli2022}. For the case of SFWM, the ring is pumped at one of its resonant frequencies $\omega_P$, and photon pairs are generated 
symmetrically around $\omega_P$ in a comb of frequency bins separated by the
free spectral range (FSR) leading to the generation of high-dimensional entangled states \cite{LiscidiniHD2019,Imany2018,KuesHD2017,ReimerComb2016}.

The simultaneous entanglement of multiple DOFs is called hyperentanglement. The original proposal of hyperentangled states was made by Kwiat (1997) \cite{Kwiat1997} and was later demonstrated by Barreiro \textit{et al.} (2005) \cite{Barreiro2005}, where they verified the simultaneous entanglement of the spatial, polarization, and time DOFs of pairs of photons generated by SPDC in a nonlinear crystal. More recently,  sources of polarization-frequency \cite{Lu2023} and polarization-mode \cite{Chiriano2023} hyperentangled photons have been demonstrated using periodically-poled lithium niobate (PPLN) crystals \cite{Lu2023} and aperiodically-poled potassium titanyl phosphate (KTP) crystals \cite{Chiriano2023} in a Sagnac loop. On-chip sources of hyperentanglement have also been demonstrated, such as polarization-frequency entangled photon pairs generated using a Bragg microcavity made from alternating layers of AlGaAs and aluminum\cite{Francesconi2023} or generated in semiconductor waveguides \cite{Dongpeng2014}, and polarization-path entangled photons, where polarization entangled photon pairs are generated via SPDC in a nonlinear crystal and their path DOFs are entangled on-chip using a beamsplitter \cite{CiampiniHE2016}.
 
Hyperentangled states carry more information than single DOF entangled states, improving the channel capacity and noise resiliency in quantum communication \cite{Barreiro2008HE, Nemirovsky-Levy2024, Kim2021,EckerQKD2019,Hu2021HD}, and increasing the key capacity and security in entanglement-based quantum key distribution (QKD) \cite{Zhong2015HD, Lana2010,Islam2017}. It has been demonstrated that hyperentangled states enable deterministic quantum information processing (QIP), originally for few qubit operations \cite{Fiorentino2004}, but recently for operations involving high-dimensional photon states \cite{Imany2019}.

Unlike previous demonstrations of polarization-frequency hyperentangled photon pairs \cite{Lu2023,Francesconi2023} where one or both DOFs were entangled off-chip, we propose an integrated source of polarization-frequency hyperentangled photon pairs in which both DOFs are simultaneously entangled on-chip. This reduces the overall losses that one would incur from coupling an off-chip source of frequency entanglement to the chip using an optical fiber \cite{Francesconi2023}. On-chip sources of polarization-frequency entangled photon pairs have been previously proposed in semiconductor waveguides \cite{Dongpeng2014}.
Our strategy uses a series of four ring resonators instead. In each ring a pump field generates a frequency-entangled signal and idler photon pair by SFWM in two different frequency bins, where both signal and idler photons are created with the same polarization, either horizontal ($H$) or vertical ($V$). The total generated state from all the rings is a coherent superposition of the two-photon states from each ring. Through engineering the ring resonators we can control the relative positions of the frequency-bins and the polarization of the generated photons in each bin. We show that this leads to a polarization-frequency hyperentangled state for the system. Finally, by manipulating the phase and intensity of the pump incident on each ring we can generate different hyperentangled states that are products of general Bell states for the polarization and frequency DOFs. The benefit of using ring resonator sources rather than waveguide, PPLN, or microcavity sources \cite{Lu2023,Francesconi2023, Dongpeng2014}, is that the dimension of the frequency DOF can be easily increased by utilizing more than two frequency-bins  in each ring.

\section{System for generating hyperentangled photon pairs}
\label{sec:system description}
\begin{figure}[htb]
    \centering
    \includegraphics[width=0.47\textwidth]{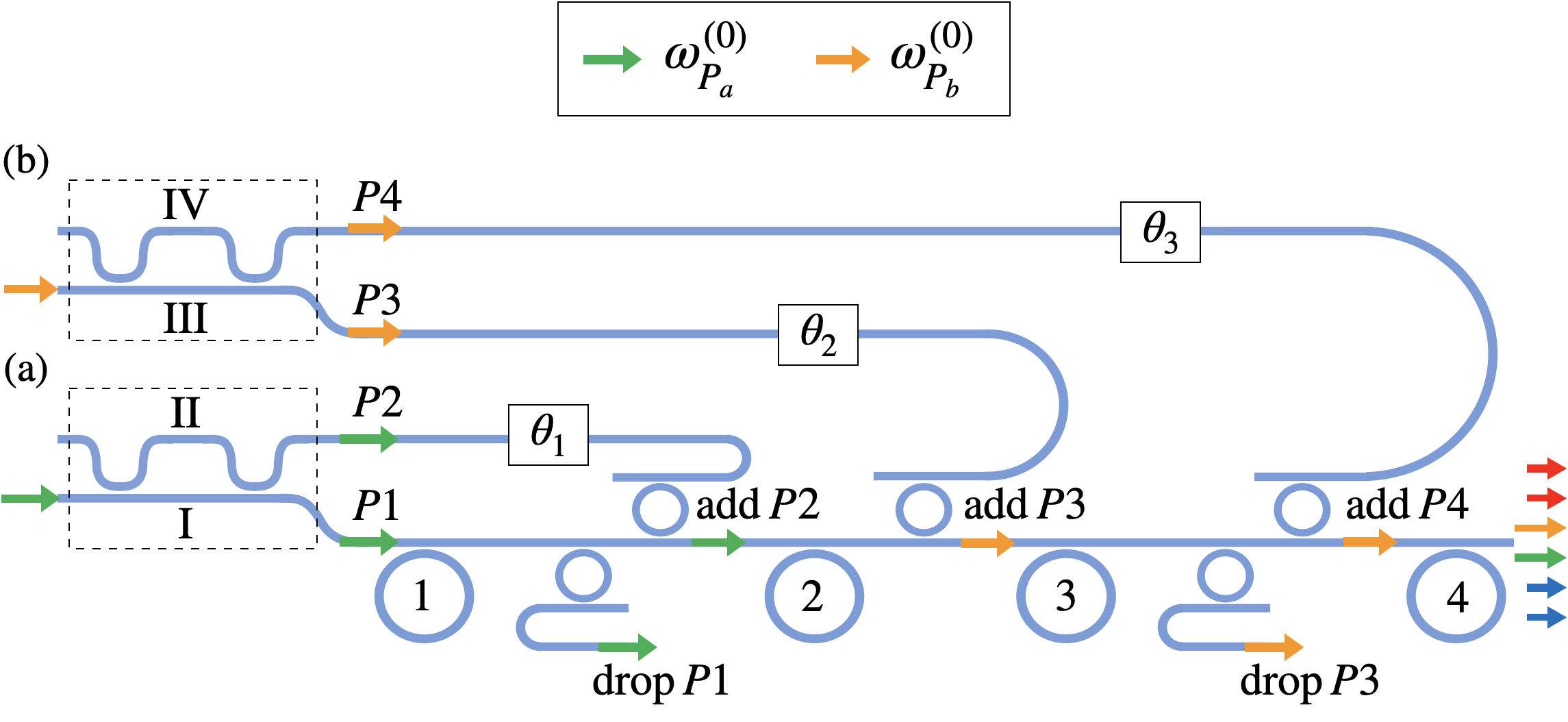}
    \caption{ Schematic of the system used to generate polarization-frequency hyperentangled photon pairs. All structures are made from SiN waveguides fully clad in SiO$_2$.}
    \label{fig:system}
\end{figure}

We consider the system shown in Fig. \ref{fig:system}. This is similar to the one previously introduced by  Liscidini and  Sipe \cite{LiscidiniHD2019} for the generation of high-dimensional entangled photon pairs, but here we extend that work to incorporate hyperentanglement in polarization and frequency DOFs. The design involves four ring resonators in series (labelled 1, 2, 3, 4), which are point-coupled to a waveguide. Pump fields are directed to each ring by a sequence of add-drop filters, and the incident  pump to each ring has its power and phase controlled by   integrated tunable Mach-Zehnder interferometers (MZIs) \cite{castro2022}. We assume all the structures are made from silicon nitride (SiN) waveguides fully clad in silicon dioxide (SiO$_2$).  Before we discuss the generation of  photon pairs we explain the propagation of the pump fields in our system. The control of the pump power and phase incident on each ring is crucial to the generation of the hyperentangled state, as we will show below.

 There are four waveguides in our system that propagate the pump fields, which are labelled with  Roman numerals I, II, III, and IV in Fig. \ref{fig:system}. These waveguides extend over the whole structure, each with an input on the leftmost side of the diagram. In the input of waveguide I we inject a pump field with center frequency $\omega^{(0)}_{P_a} =2\pi \times 193.415\,{\rm THz}$ (green arrow), and in the input region of waveguide III we inject a pump field with center frequency $\omega^{(0)}_{P_b} =2\pi \times 192.175\,{\rm THz}$  (orange arrow). No fields are injected in the remaining waveguides labelled II and IV . Each waveguide I, II, III, and IV has the same thickness and width of $800\,{\rm nm}$, making a square cross-section, which is uniform everywhere in the structure. For our frequencies of interest this guarantees that each waveguide  approximately supports only the fundamental TE mode  and fundamental TM mode, allowing us to neglect the higher-order TE and TM modes of the waveguide, and  so that the fundamental TE and TM  modes have  approximately the same effective index. For the remainder of this paper we use the convention that a photon in the fundamental TE mode of a waveguide has horizontal ($H$) polarization, and a photon in the fundamental TM mode of a waveguide has vertical ($V$) polarization.

In Fig. \ref{fig:system} the pumps injected into waveguides I and III are each in a superposition of $H$ and $V$ polarization. The two MZIs of our system are identified by the dashed boxes in Fig. \ref{fig:system}.  MZI (a)  takes the incident pump (green arrow) in waveguide I and couples it to waveguide II using directional couplers (DCs) with a path imbalance. The two exiting pumps $P1$ and $P2$ (green arrows) are each in superposition of $H$ and $V$ polarization. The relative phase and power between $P1$ and $P2$ can be adjusted using a thermal component in between the two DCs on  waveguide II \cite{castro2022}.  The MZI (b) operates similarly, such that it adjusts the relative phase and power in the pump fields $P3$ and $P4$ (orange arrows), which are each in a superposition of $H$ and $V$ polarization. 

After MZI (a) $P1$ is directed to ring 1. Its accumulated phase before it enters ring 1 is due to MZI (a) and the distance it travelled. The pump $P2$ is added to waveguide I using the critically coupled add filter indicated in Fig. \ref{fig:system}. The accumulated phase of $P2$ before it enters ring 2 is due to the MZI (a), the distance it travelled, and a $\pi$-phase from the add filter. The total relative phase between the field incident on ring 2 and that incident on ring 1 is denoted by $\theta_1$ in Fig. \ref{fig:system}. Similarly, pumps $P3$ and $P4$ are added to waveguide I using critically coupled add filters, and directed to rings 3 and 4 as indicated in Fig. \ref{fig:system}. We denote by $\theta_2$ the total relative phase  between the fields incident on rings 3 and 1, and  by $\theta_3$ the total relative phase  between the fields incident on rings 4 and  1. The relative phases $\theta_1$, $\theta_2$, and $\theta_3$ can be controlled with heaters placed on waveguides II, III, and IV after each MZI.

The four labelled ring resonators in Fig. \ref{fig:system} are engineered so that ring 1 only accepts the $H$ polarization of $P1$, allowing the $V$ polarization of $P1$ to pass. Ring 2 has a resonance for the $V$ polarization of both $P1$ and $P2$, but the critically coupled drop filter after ring 1 removes both polarizations of $P1$ to guarantee that ring 2 only contains the $V$ polarization of $P2$. Similarly, ring 3 only accepts the $H$ polarization of $P3$  and the critically coupled drop filter after ring 3 guarantees that ring 4 only contains the $V$ polarization of $P4$. No drop filter for $P2$ is required since it is off-resonance with rings 3 and 4.  We refer to the above constraints, where each ring accepts only a specific pump frequency and polarization, as the tuning conditions.

\begin{figure}[htb]
    \centering
    \includegraphics[width=0.37\textwidth]{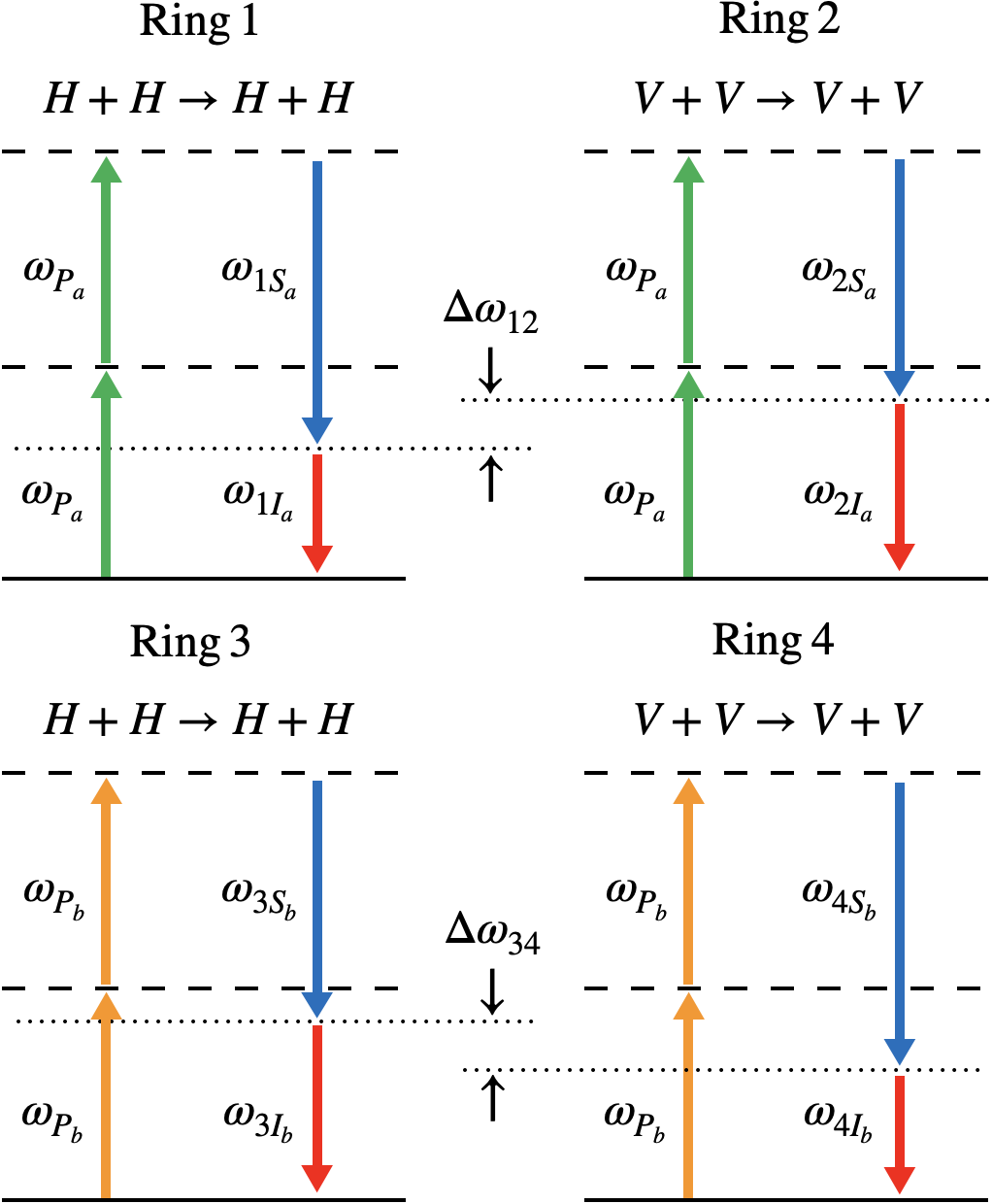}
    \caption{Energy diagrams for the SFWM interactions in rings 1, 2, 3, and 4.}
    \label{fig:energy diagram}
\end{figure}

In each ring two pump photons can be destroyed and  a signal and idler photon created by a SFWM interaction that conserves energy. The energy diagrams for the interactions in the rings 1, 2, 3 and 4 are shown in Fig. \ref{fig:energy diagram}. We consider three resonances of each ring.
 Rings 1 and 2 share the pump resonance $P_a$, but ring 1 has the signal and idler resonances $1S_a$ and $1I_a$, while 
 for ring 2 they are $2S_a$ and $2I_a$. Similar processes occur in rings 3 and 4 as indicated in Fig. \ref{fig:energy diagram}, except these rings share the common resonance $P_b$ instead, and the associated signal and idler resonances are given by $3S_b$, $3I_b$, $4S_b$, and $4I_b$. 
 
In general the signal photons from rings 1 and 2 can be separated in frequency by the frequency difference $\Delta \omega_{12}$ as indicated in Fig. \ref{fig:energy diagram}, which is caused by unequal FSRs. To generate polarization-frequency hyperentanglement a necessary requirement is that the signal photons from rings 1 and 2, which have opposite polarizations, are created in the same frequency-bin. When the difference in the FSRs of rings 1 and 2 are much less than a resonance linewidth we expect that $\Delta \omega_{12} \approx 0$. Using similar arguments we expect that $\Delta \omega_{34} \approx 0$ when the FSRs of rings 3 and 4 are close. It was recently demonstrated in a similar system that two resonances from different rings can be spectrally aligned so that they share a single transmission dip \cite{Clementi2023}.

\begin{figure}[htb]
    \centering
    \includegraphics[width=0.41\textwidth]{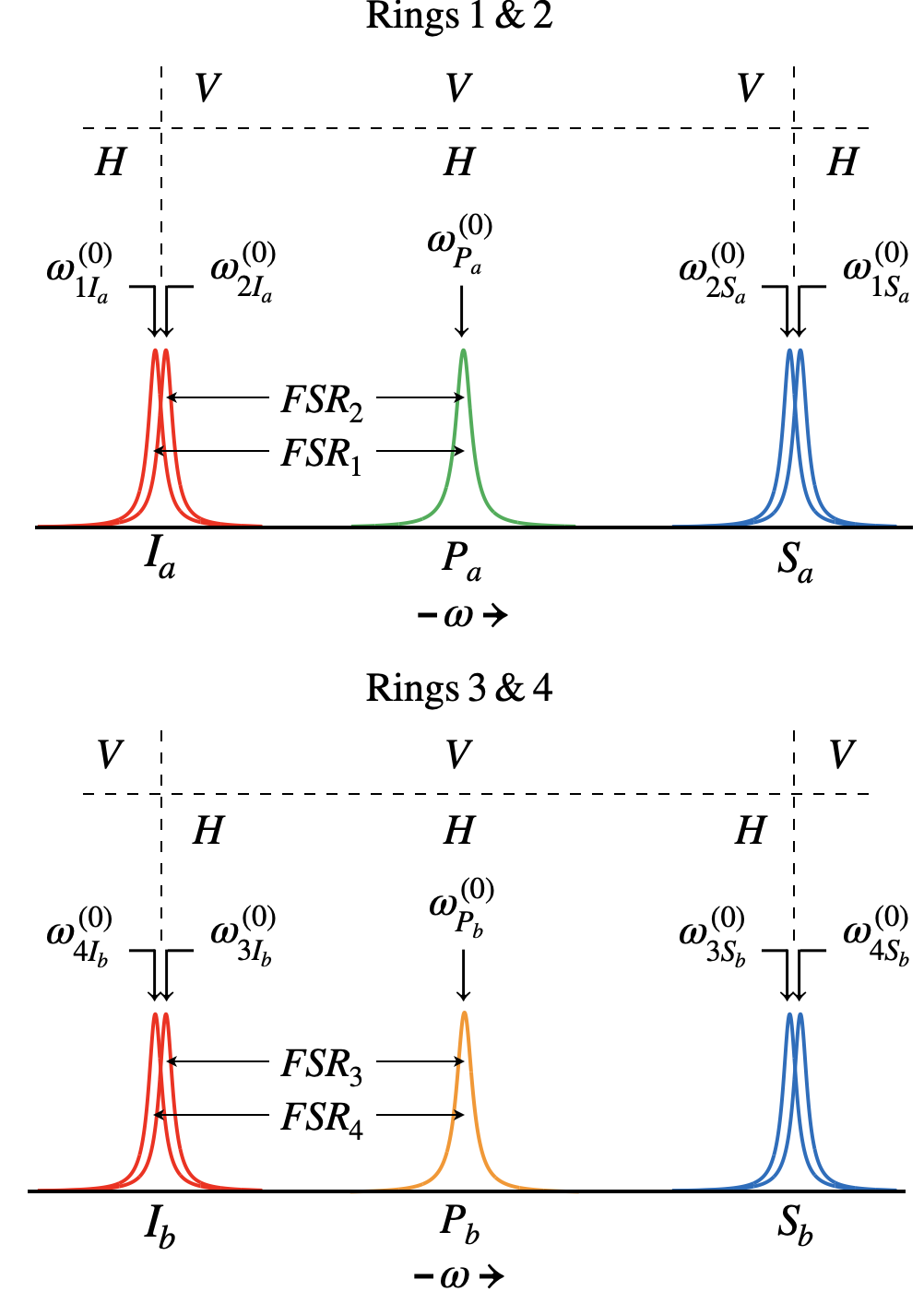}
    \caption{Schematic representation of the positions of the ring resonances for the four rings. The center frequencies of the resonances for the $H$ and $V$ polarization and the FSRs of the rings are labelled. The differences between the FSRs are greatly exaggerated. We also introduce the signal and idler frequency bins $S_a, \,I_a$ associated with pump $P_a$, and bins $S_b, \,I_b$ associated with pump $P_b$.  }
    \label{fig:resonances}
\end{figure}

In our simulations we achieve $\Delta \omega_{12} \approx 0$ by varying the widths of rings 1 and 2 until their FSRs are approximately equal. We achieve $\Delta \omega_{34}\approx 0$ similarly by changing the widths of rings 3 and 4. Choosing the widths of rings 1, 2, 3, and 4 to be $W_1 = W_3 = 900\,{\rm nm}$ and $W_2 = W_4 = 825\,{\rm nm}$, we numerically calculate a difference in the FSRs of rings 1 and 2 of $|FSR_1 - FSR_2| = 2\pi \times 0.3\,{\rm GHz}$ and virtually the same for rings 3 and 4. In Fig. \ref{fig:resonances} a schematic representation of the positions of the ring resonances for our system are shown, where the center frequencies of the resonances, the polarization of the photons in each resonance, and the FSRs are indicated, where $FSR_1 = 283.9\,{\rm GHz}$ and $FSR_2 = 283.6\,{\rm GHz}$. Assuming a loaded quality factor for each resonance of $Q_{\rm load} = 5\times 10^5$, the differences in the FSRs are approximately 4 times smaller than a typical linewidth of $2\pi\times 1.2\,{\rm GHz}$. The small differences in the FSRs are illustrated in Fig. \ref{fig:resonances} by overlapping signal and idler resonances; however, the differences are greatly exaggerated in that figure for illustrative purpose.

For the remainder of this paper we assume that differences in the FSRs are sufficiently smaller than the linewidths, such that our system effectively only contains the resonances $S_a$, $I_a$, $S_b$, and $I_b$ as indicated in Fig. \ref{fig:resonances}. As a result of this, the signal and idler photons from rings 1 and  2  are generated in $S_a$ and $I_a$, respectively, while the signal and idler photons from rings 3 and 4 are generated in $S_b$ and $I_b$ respectively.

\section{Generated state of photon pairs}
Having described the system, we now write down the total state that the system generates and demonstrate in the next section that it describes polarization-frequency hyperentangled photon pairs. Taking the limit of a small generation probability, we assume that at most one 
photon pair is generated in the structure. The output state for the photons can be written as a superposition of the vacuum state and a two-photon state from each ring \cite{OnoderaRings2016}:
\begin{align}
    \label{eq:4 ring state}
    \ket{\psi} &= \frac{1}{|\beta|}\sum_{n=1}^4 \beta_n\ket{\psi_n},
\end{align}
where $\ket{\psi}$ is the state of the system given that a pair of photons is generated, $\beta_n$ is the amplitude for the two-photon state $\ket{\psi_n}$ for ring $n$, and $|\beta| \equiv (\sum_{n=1}^4|\beta_n|^2)^{1/2}$ normalizes the total state. Under our assumption that at most a photon-pair is created, $|\beta|^2 \ll 1$ is the probability of generating a pair per pulse. The two-photon states for each ring are written as
\begin{align}
\label{eq:ring 1 ket}
    \ket{\psi_1} &= \int d\Omega d\Omega' \phi_{1} (\Omega, \Omega') \ket{ H,S_a;\Omega}\ket{ H,I_a;\Omega'}, 
    \\
    \label{eq:ring 2 ket}
    \ket{\psi_2} &= \int d\Omega d\Omega' \phi_{2} (\Omega, \Omega') \ket{ V,S_a;\Omega}\ket{ V,I_a;\Omega'},
      \\
    \label{eq:ring 3 ket}
    \ket{\psi_3} &= \int d\Omega d\Omega' \phi_{3} (\Omega, \Omega') \ket{ H,S_b;\Omega}\ket{ H,I_b;\Omega'},
          \\
    \label{eq:ring 4 ket}
     \ket{\psi_4} &= \int d\Omega d\Omega' \phi_{4} (\Omega, \Omega') \ket{ V,S_b;\Omega}\ket{ V,I_b;\Omega'},
\end{align}
where $\phi_{n}(\Omega, \Omega')$ is the normalized biphoton wave function for ring $n$, as shown in Appendix A
,and we have defined the composite kets as
\begin{align}
\label{eq:composite ket}
\ket{P, B; \Omega} &\equiv a^\dagger_{PB}(\Omega)\ket{\rm vac},
\end{align}
where  $ a^\dagger_{PB}(\Omega) $  creates a photon with frequency detuning $\Omega = \omega - \omega^{(0)}_{nB}$, polarization $P=H,V$, and in  frequency-bin $B = S_a,\,I_a,\,S_b,\,I_b$. The creation and annihilation operators satisfy the commutation relations
\begin{align}
\label{eq:commute a}
    [a_{PB}(\Omega),a^\dagger_{PB}(\Omega') ] &= \delta(\Omega - \Omega'),
\end{align}
 and all others are zero. 

 Given that a pair of photons is generated, the probability for each ring is $|\beta_n|^2$. 
 Using the pump MZIs  (see Fig. \ref{fig:system}) one can adjust the pump power incident on each ring such that $|\beta_n|^2$ is the same for each ring. We put
 \begin{align}
 \label{eq:equal power}
     |\beta_1|^2 = |\beta_2|^2 =|\beta_3|^2 = |\beta_4|^2 =\frac{1}{4}|\beta|^2,
 \end{align}
 such that the relative probability for pair to be generated in a given ring is $1/4$. The magnitudes $|\beta_n|$ are set by Eq. \eqref{eq:equal power}, but the phases of each amplitude still have to be set. We have the freedom to choose the phase of the  $\beta_n$ for each ring so that it cancels the relative phases indicated in Fig. \ref{fig:system}. This results in
 \begin{align}
 \label{eq:betas}
     \beta_2 = \beta_1 {\rm e}^{2i\theta_1}, \,\,\,\,\,
\beta_3 = \beta_1 {\rm e}^{2i\theta_2},\,\,\,\,\, \beta_4 = \beta_1 {\rm e}^{2i\theta_3},
 \end{align}
 where the factor of $2$ arises from the biphoton wavefunction for SFWM, as shown in Appendix A
 , being proportional to the pump amplitude squared \cite{MBanicRingLoss2022}.

\section{Demonstration of hyperentanglement}
The demonstration of hyperentanglement can be carried out by showing that the photons are entangled in both the polarization and frequency-bin subspaces, when considered separately, and that the states in the two DOFs are uncorrelated. We do this by tracing over either the polarization or frequency-bin DOF of the ket in Eq. \eqref{eq:4 ring state} and showing that the resulting state is pure and entangled in the other DOF. To perform the trace over the individual DOFs of the photon pairs in the state in Eq. \eqref{eq:4 ring state}, we need to separate the photon's polarization, frequency-bin, and continuous-frequency detuning DOFs. 

One way to identify the DOFs is by defining the quantum Stokes operators \cite{Schlichtholz2022}, which form a set of compatible observables that can be used to measure the total photon number and obtain information about the photon polarization. We define the four Stokes operators for each frequency-bin $B$ as
\begin{align}
\label{eq:S0}
    \Sigma_{0B}&=  \int d \Omega \left(a^\dagger_{HB}(\Omega)a_{HB}(\Omega) + a^\dagger_{VB}(\Omega)a_{VB}(\Omega)\right),
    \\
    \label{eq:S1}
     \Sigma_{1B} &=\int d \Omega \left(a^\dagger_{HB}(\Omega)a_{HB}(\Omega) - a^\dagger_{VB}(\Omega)a_{VB}(\Omega)\right),
         \\
    \label{eq:S2}
     \Sigma_{2B} &=\int d \Omega \left(a^\dagger_{HB}(\Omega)a_{VB}(\Omega) + a^\dagger_{VB}(\Omega)a_{HB}(\Omega)\right),
              \\
    \label{eq:S3}
     \Sigma_{3B} &=-i\int d \Omega \left(a^\dagger_{HB}(\Omega)a_{VB}(\Omega) - a^\dagger_{VB}(\Omega)a_{HB}(\Omega)\right),
\end{align}
where recall $B = S_a,\,I_a,\,S_b,\,I_b$. Here $\Sigma_{0B}$ is the total photon number operator for bin $B$, $\Sigma_{1B}$ is the difference between the number of $H$ and $V$ polarization photons, $\Sigma_{2B}$ is the difference between the diagonal and anti-diagonal polarized photons, and $\Sigma_{3B}$ represents the difference between left-hand and right-hand circularly polarized photons; the diagonal and anti-diagonal  polarized states and the left-hand and right-hand circularly polarized states are coherent superpositions of the states identified by $H$ and $V$. For a given bin $B$, \{$\Sigma_{iB}$\} are associated with the usual Stokes parameter introduced to describe light polarization. It follows that the polarization operator $\hat{P}$ can be constructed as
\begin{align}
    \label{eq:polarization operator}
    \hat{P} = \sum_B \Sigma_{1B},
\end{align}
where the sum is over all the frequency bins. Putting Eq. \eqref{eq:S1} for $\Sigma_{1B}$ into Eq. \eqref{eq:polarization operator}, the operator $\hat{P}$ is the total photon number operator for $V$ polarization subtracted from the total photon number operator for $H$ polarization. So $\hat{P}$ describes the total amount of $H$ polarization relative to the total amount of $V$ polarization. Finally, we can define frequency detuning operator
\begin{align}
\label{eq:freq operator}
    \hat{\Omega} = \int d\Omega \, \Omega  \hat{N}(\Omega),
\end{align}
where we have defined the total photon number operator for the frequency detuning $\Omega$ as
\begin{align}
    \hat{N}(\Omega) \equiv \sum_{B}\sum_{P} a^\dagger_{PB}(\Omega) a_{PB}(\Omega),
\end{align}
obtained by summing over the all frequency-bins $B$ and polarizations $P = H,V$.

The operators $\Sigma_{0B}$, $\hat{P}$, and $\hat{\Omega}$ describe the frequency-bin, polarization, and continuous-frequency detuning DOFs of the photon. These operators commute with each other
\begin{align}
    \label{eq:commute 3 ops}
    \left[\Sigma_{0B}, \hat{P}\right] = \left[\Sigma_{0B}, \hat{\Omega}\right] = \left[\hat{P}, \hat{\Omega}\right] = 0,
\end{align}
where we used Eq. \eqref{eq:commute a} for the commutation relations of the photon operators. Because $\Sigma_{0B}$, $\hat{P}$, and $\hat{\Omega}$ mutually commute we can find a set of common eigenkets. It is easy to confirm that the set of common  eigenkets is given by Eq. \eqref{eq:composite ket} for the composite kets $\ket{P,B;\Omega}$, where the eigenvalues of $\Sigma_{0B}$ are $0$ and $1$, the eigenvalues of $\hat{P}$ are $+1$ for $H$ polarization and $-1$ for $V$ polarization, and the eigenvalues of $\hat{\Omega}$ are $\Omega$.

Since the operators $\Sigma_{0B}$, $\hat{P}$, and $\hat{\Omega}$ each describe  a DOF,  we decompose the total Hilbert space $\mathcal{H}_{\rm photon}$ for a single photon into  three Hilbert spaces, each corresponding to a different DOF of the photon
\begin{align}
    \mathcal{H}_{\rm photon} =   \mathcal{H}_{\rm pol}\otimes \mathcal{H}_{\rm bin} \otimes\mathcal{H}_{\rm detun},
\end{align}
where $\mathcal{H}_{\rm pol}$, $\mathcal{H}_{\rm bin}$, and $\mathcal{H}_{\rm detun}$ are the Hilbert spaces for the photon's frequency-bin, polarization, and continuous-frequency detuning DOFs, respectively. We introduce the frequency-bin kets $\ket{B}$, polarization kets $\ket{P}$, continuous-frequency detuning kets $\ket{\Omega}$, which span the Hilbert spaces $\mathcal{H}_{\rm bin}$,  $\mathcal{H}_{\rm pol}$, and  $\mathcal{H}_{\rm detun}$, respectively,  where their inner products are defined as
\begin{align}
     \label{eq:inner prod pol}
    \bra{P}\ket{P'} &= \delta_{PP'},
    \\
     \label{eq:inner prod bin}
    \bra{B}\ket{B'} &= \delta_{BB'},
    \\
    \label{eq:inner prod freq}
    \bra{\Omega}\ket{\Omega'} &= \delta(\Omega-\Omega'),
\end{align}
and the identity operators for each Hilbert space are given by
\begin{align}
    \hat{\mathbb{1}}_{\rm pol} &= \sum_{P} \ket{P}\bra{P},
    \\
        \hat{\mathbb{1}}_{\rm bin} &= \sum_{B} \ket{B}\bra{B},
    \\
     \hat{\mathbb{1}}_{\rm detun} &= \int d\Omega \ket{\Omega}\bra{\Omega}.
\end{align}
Since $\Sigma_{0B}$, $\hat{P}$, and $\hat{\Omega}$  share a common set of eigenkets $\ket{P, B;\Omega}$ the action of each operator on the eigenkets only affect a single DOF  and leave the remaining two DOFs unchanged. Thus we can think of each operator as acting non-trivially only over the Hilbert space for one DOF and acting with the identity operators over the Hilbert spaces for the remaining two DOFs,
\begin{align}
  \label{eq:pol decompose}
    \hat{P}&\rightarrow  \hat{P}^{(\rm pol)}\otimes \hat{\mathbb{1}}_{\rm bin} \otimes \hat{\mathbb{1}}_{\rm detun} ,
        \\
        \label{eq:bin decompose}
    \Sigma_{0B}&\rightarrow \hat{\mathbb{1}}_{\rm pol} \otimes \Sigma^{(\rm bin)}_{0B}\otimes \hat{\mathbb{1}}_{\rm detun} ,
    \\
  \label{eq:freq decompose}
    \hat{\Omega}&\rightarrow \hat{\mathbb{1}}_{\rm pol} \otimes  \hat{\mathbb{1}}_{\rm bin} \otimes \hat{\Omega}^{(\rm detun)},
\end{align}
where we have introduced the operators $\Sigma^{(\rm bin)}_{0B}$, $\hat{P}^{(\rm pol)}$, and $\hat{\Omega}^{(\rm detun)}$ that only act over the Hilbert space for a single DOF. The operators $\Sigma^{(\rm bin)}_{0B}$, $\hat{P}^{(\rm pol)}$, and $\hat{\Omega}^{(\rm detun)}$  satisfy the eigenvalue equations
\begin{align}
    \label{eq:eig pol H}
     \hat{P}^{(\rm pol)} \ket{H} &=  \ket{H},
         \\
         \label{eq:eig pol V}
     \hat{P}^{(\rm pol)} \ket{V} &=  -\ket{V},
              \\
              \label{eq:eig bin}
    \Sigma^{(\rm bin)}_{0B} \ket{B'} &= \delta_{BB'} \ket{B'},
    \\
              \label{eq:eig freq}
     \hat{\Omega}^{(\rm detun)} \ket{\Omega} &=  \Omega\ket{\Omega}.
\end{align}

Now consider the action of the identity operator $\hat{\mathbb{1}}$ on Eq. \eqref{eq:composite ket} for the eigenkets
\begin{align}
\label{eq:eigenket expand}
 \hat{\mathbb{1}}\ket{P,B;\Omega} &= ( \hat{\mathbb{1}}_{\rm pol}\otimes \hat{\mathbb{1}}_{\rm bin}\otimes\hat{\mathbb{1}}_{\rm detun} )\ket{P,B;\Omega}\nonumber
    \\
    &=\sum_{B'}\sum_{P'}\int d\Omega' c_{BPB'P'}(\Omega, \Omega') \nonumber
    \\
    &\times\ket{P'}\otimes \ket{B'}\otimes\ket{\Omega'},
\end{align}
where we have defined the coefficients
\begin{align}
\label{eq:coefficients}
    c_{BPB'P'}(\Omega, \Omega') &\equiv \bra{\Omega'} \otimes \bra{B'} \otimes \bra{P'} \ket{P,B;\Omega}.
\end{align}
We show in Appendix B
that the coefficients are given by
\begin{align}
\label{eq:coefficients solution}
    c_{BPB'P'}(\Omega, \Omega') = \delta_{BB'}\delta_{PP'}\delta(\Omega-\Omega'),
\end{align}
where we have neglected a multiplicative phase factor in Eq. \eqref{eq:coefficients solution} that depends on $P$, $B$, and $\Omega$.
Putting Eq. \eqref{eq:coefficients solution} into Eq. \eqref{eq:eigenket expand} we obtain
\begin{align}
    \label{eq:ket separated}
    \ket{P,B;\Omega} = \ket{P}\otimes\ket{B}\otimes\ket{\Omega},
\end{align}
which is the desired separation of the composite ket into a product of three kets, one for each DOF.

We are now in a position to write Eq. \eqref{eq:4 ring state} for the total state in a form where the DOFs of the individual photons are separated. To do this we put Eq. \eqref{eq:ket separated} for the decomposed composite kets into Eq. \eqref{eq:ring 1 ket} -- Eq. \eqref{eq:ring 4 ket} for the two-photon state for each ring; then putting these into Eq. \eqref{eq:4 ring state}, we obtain
\begin{align}
\label{eq:4 ring state 2}
    \ket{\psi} &= \frac{1}{2} \Big( \ket{HH}\ket{S_aI_a}\ket{11} +  {\rm e}^{2i\theta_1} \ket{VV} \ket{S_aI_a} \ket{22}\nonumber
    \\
    &+ {\rm e}^{2i\theta_2}\ket{HH} \ket{S_bI_b}\ket{33}+{\rm e}^{2i\theta_3}\ket{VV}\ket{S_bI_b}\ket{44} \Big),
\end{align}
where we have used Eq. \eqref{eq:equal power} and Eq. \eqref{eq:betas} for the magnitude and phase of each amplitude. The kets $\ket{HH} = \ket{H}\otimes \ket{H}$ and $\ket{VV} = \ket{V}\otimes \ket{V}$ correspond to both the signal and idler photon having $H$ and $V$ polarization, and the kets $\ket{S_a I_a}  = \ket{S_a} \otimes \ket{I_a}$  and $\ket{S_b I_b}  = \ket{S_b} \otimes \ket{I_b}$ corresponds to a signal photon in the bin $S_a$ and idler in the bin $I_a$, and a signal photon in the bin $S_b$ and idler in the bin $I_b$. We have also defined the ket $\ket{nn}$ for ring $n$ as
\begin{align}
\label{eq:ring n ket}
    \ket{nn} \equiv \int d\Omega d\Omega' \phi_n (\Omega, \Omega') \ket{\Omega}\otimes\ket{\Omega'},
\end{align}
where $\ket{\Omega}$ is for the signal  photon and $\ket{\Omega'}$ is for the idler photon.

The density operator for the system is given by $\rho = \ket{\psi}\bra{\psi}$, where $\ket{\psi}$ is given by Eq. \eqref{eq:4 ring state 2}. We show in Appendix C
that by tracing over the continuous-frequency detuning DOF $\Omega$ and $\Omega'$ of $\rho$ we obtain the reduced density operator, $\overline{\rho} = {\rm Tr}_{\Omega \Omega'}[\rho]$, for the polarization and frequency-bin DOFs only. Taking $\overline{\rho}$ and tracing over the frequency-bin DOF, we obtain the reduced density operator for the polarization DOF only, $\overline{\rho}_{\rm pol} = {\rm Tr}_{\rm bin}[\overline{\rho}]$, which can be written as
\begin{align}
    \label{eq:rho HV}
   \overline{\rho}_{\rm pol} &=\frac{1}{2}\left(\ket{HH}\bra{HH} +  \ket{VV}\bra{VV}\right) \nonumber
   \\
   &+ \frac{1}{4}\left({\rm e}^{2i\theta_1}O^*_{12} + {\rm e}^{2i(\theta_3-\theta_2)}  O^*_{34}\right)\ket{VV}\bra{HH}\nonumber
   \\
   &+\frac{1}{4}\left({\rm e}^{-2i\theta_1}O_{12} + {\rm e}^{-2i(\theta_3-\theta_2)}  O_{34}\right)\ket{HH}\bra{VV},
\end{align}
 where we have defined the overlap integral of the biphoton wavefunctions from rings $n$ and $n'$ as
\begin{align}
\label{eq:overlaps}
    O_{nn'}\equiv \int d\Omega d\Omega' \phi_n(\Omega,\Omega')\phi^*_{n'}(\Omega,\Omega'),
\end{align}
where $O^*_{n'n} = O_{nn'}$. Finally, taking $\overline{\rho}$ and tracing over the polarization DOF, we obtain the reduced density operator for the frequency-bin DOF only, $\overline{\rho}_{\rm bin} = {\rm Tr}_{\rm pol}[\overline{\rho}]$, which can be written as
\begin{align}
    \label{eq:rho SI}
   \overline{\rho}_{\rm bin} &=\frac{1}{2}\left(\ket{S_aI_a}\bra{S_aI_a} +  \ket{S_bI_b}\bra{S_bI_b}\right) \nonumber
   \\
   &+ \frac{1}{4}\left({\rm e}^{2i\theta_2}O^*_{13} + {\rm e}^{2i(\theta_3-\theta_1)}O^*_{24}\right)\ket{S_bI_b}\bra{S_aI_a}\nonumber
   \\
   &+ \frac{1}{4}\left({\rm e}^{-2i\theta_2}O_{13} + {\rm e}^{-2i(\theta_3-\theta_1)}O_{24}\right)\ket{S_aI_a}\bra{S_bI_b}.
\end{align}
The reduced density operators in Eq. \eqref{eq:rho HV} and Eq. \eqref{eq:rho SI} for the polarization DOF and frequency-bin DOF generally correspond to mixed states in either DOF. However, to demonstrate hyperentanglement between the polarization and frequency-bin DOFs,  we require that the reduced density operators correspond to pure entangled states in either DOF. 

We start by calculating the purity of the reduced density operator for the polarization DOF, which is given by $\gamma_{\rm pol}={\rm Tr}(\overline{\rho}^2_{\rm pol})$. By using Eq. \eqref{eq:rho HV} for $\overline{\rho}_{\rm pol}$, one obtains
\begin{align}
\label{eq:purity HV}
   \gamma_{\rm pol} &= \frac{1}{2} + \frac{1}{8}\Big(|O_{12}|^2 + |O_{34}|^2 \nonumber
   \\
   &+ 2\Re{{\rm e}^{2i(\theta_3 - \theta_2 - \theta_1)} O_{12}O^*_{34}}\Big).
\end{align}

To have unit purity, $\gamma_{\rm pol} =1$,  two conditions must be satisfied. First, the overlap integrals (\ref{eq:overlaps}) must be unity, $O_{nn'} = 1$, which will be the case if the biphoton wavefunctions $\phi_n(\Omega,\Omega^\prime)$ describing the spectral detuning of the generated photons are the same for each ring $n$.  The second condition to be satisfied is $\cos(2(\theta_3 -\theta_2 - \theta_1)) = 1$, which requires that the phases be set carefully to achieve hyperentanglement. This condition arises from considering four different rings where each ring generates a separable state in terms of the polarization and frequency-bin DOFs. This condition is not present when working with single DOF entanglement \cite{Clementi2023}. We can satisfy this second condition by adjusting $\theta_3$ to  $\theta_3 = \theta_1+\theta_2 + \pi m$, for any integer $m$. 
Assuming that these two conditions are satisfied by the system, Eq. \eqref{eq:rho HV} for $\overline{\rho}_{\rm pol}$ can be written as
\begin{align}
    \label{eq:rho HV unit purity}
   \overline{\rho}_{\rm pol}^{\rm pure} &=\frac{1}{2}\left(\ket{HH}+  {\rm e}^{2i\theta_1}\ket{VV}\right)\nonumber
   \\
   &\otimes \left(\bra{HH}+  {\rm e}^{-2i\theta_1}\bra{VV}\right), 
\end{align}
which is the density operator for the Bell state $(\ket{HH}+  {\rm e}^{2i\theta_1}\ket{VV})/\sqrt{2}$. Note that in our setup in Fig. \ref{fig:system} one can control the phase $\theta_1$, which allow us to generate either $\Phi^+$ or $\Phi^-$, by setting $\theta_1$ equal to $\pi$ or $\pi/2$, respectively. 

Following similar steps as above, the purity of the reduced density operator for the frequency-bin DOF, $\gamma_{\rm bin}={\rm Tr}(\overline{\rho}^2_{\rm bin})$, is given by
\begin{align}
\label{eq:purity SI}
   \gamma_{\rm bin} &= \frac{1}{2} + \frac{1}{8}\Big(|O_{13}|^2 + |O_{24}|^2 \nonumber
   \\
   &+ 2\Re{{\rm e}^{2i(\theta_3 - \theta_2 - \theta_1)} O_{13}O^*_{24}}\Big),
\end{align}
where we used Eq. \eqref{eq:rho SI} for $\overline{\rho}_{\rm bin}$. To have unit purity, $\gamma_{\rm bin} =1$, the overlap integrals must be unity, $O_{nn'} = 1$ and $\theta_3 = \theta_1+\theta_2 + \pi m$, for any integer $m$. Not surprisingly, these are the same conditions derived above, for purity in one DOF requires the absence of any correlation with the others. Putting $O_{nn'} = 1$ into Eq. \eqref{eq:rho SI} for $\overline{\rho}_{\rm bin}$ we obtain 
\begin{align}
    \label{eq:rho SI unit purity}
   \overline{\rho}_{\rm bin}^{\rm pure} &=\frac{1}{2}\left(\ket{S_aI_a}+  {\rm e}^{2i\theta_2}\ket{S_bI_b}\right) \nonumber
   \\
   &\otimes \left(\bra{S_aI_a}+  {\rm e}^{-2i\theta_2}\bra{S_bI_b}\right), 
\end{align}
which is the pure density operator for the Bell state $(\ket{S_aI_a}+  {\rm e}^{2i\theta_2}\ket{S_bI_b})/\sqrt{2}$. In analogy with the case of polarization, one can control the generated state with the phase $\theta_2$ (see Fig. \ref{fig:system}). As expected, since there is no correlation between polarization and bin, the state can be set in each subspace independently.

In summary, we have demonstrated that if our system satisfies the conditions that the biphoton wavefunctions from each ring are identical, and $\theta_3 = \theta_1+\theta_2$, then the reduced density operators for the frequency-bin and polarization DOFs will have unit purity  and individually correspond to Bell states in a each DOF subspace. Assuming that these conditions are satisfied, after tracing over the continuous-frequency detuning DOF in Eq. \eqref{eq:4 ring state 2}, the reduced density operator $\overline{\rho}$ can be written as
\begin{align}
    \overline{\rho} = \ket{\psi_{\rm HE}} \bra{\psi_{\rm HE}},
\end{align}
where $\ket{\psi_{\rm HE}}$ is the polarization-frequency-bin  hyperentangled state given by
\begin{align}
    \label{eq:psi HE}
    \ket{\psi_{\rm HE}} &= \frac{1}{2}\left(\ket{HH}+  {\rm e}^{2i\theta_1}\ket{VV}\right) \nonumber
    \\
    &\otimes \left(\ket{S_aI_a}+  {\rm e}^{2i\theta_2}\ket{S_bI_b}\right).
\end{align}

\section{Conditions for hyperentanglement}
\label{sec:results HE}

As discussed in the previous section there are three conditions that need to be satisfied to obtain hyperentanglement in our system. First, the pair generation probability must be the same for each ring, which is satisfied with Eq. \eqref{eq:equal power}. Second, the relative phase $\theta_3$ should be given by $\theta_3 = \theta_1 +\theta_2$. Third, the overlap integrals of the biphoton wavefunctions (\ref{eq:overlaps}) must all equal 1. This last requirement can be challenging to achieve depending on fabrication accuracy. Below, we consider realistic parameters, evaluate the overlap integrals, and investigate how they depend on the coupling efficiencies of the rings, the quality factors of the involved resonances, and the properties of the pump pulse. 

We set the radii of rings 1 and 4 to be $R_{1} = R_4 = 80\,\mu{\rm m} - 20\,{\rm nm}$ and rings 2 and 3 to be $R_{2} = R_3 = 80\,\mu{\rm m} + 60\,{\rm nm}$ to satisfy the tuning conditions introduced in Sec. \ref{sec:system description}. The cross sections of the waveguides in the ring lead us to take the group index and effective index for the $H$ and $V$ polarization in waveguide I (which has a square cross-section) to be $ng_H = ng_V = 2.10$ and $n_{{\rm eff}H} = n_{{\rm eff}V} = 1.72$. We calculate the nonlinear parameter for SFWM \cite{MBanicRingLoss2022} in rings 1 and 3 to be $\Lambda_{1}=\Lambda_3 = 1.48\, ({\rm Wm})^{-1}$ and in rings 2 and 4 to be $\Lambda_{2}=\Lambda_4 = 1.40\, ({\rm Wm})^{-1}$. For each ring, we assume an intrinsic quality factor $Q_{\rm int} = 1\times 10^6$, uniform in the frequency range considered here.

\begin{figure}[htb]
    \centering
    \includegraphics[width=1\linewidth]{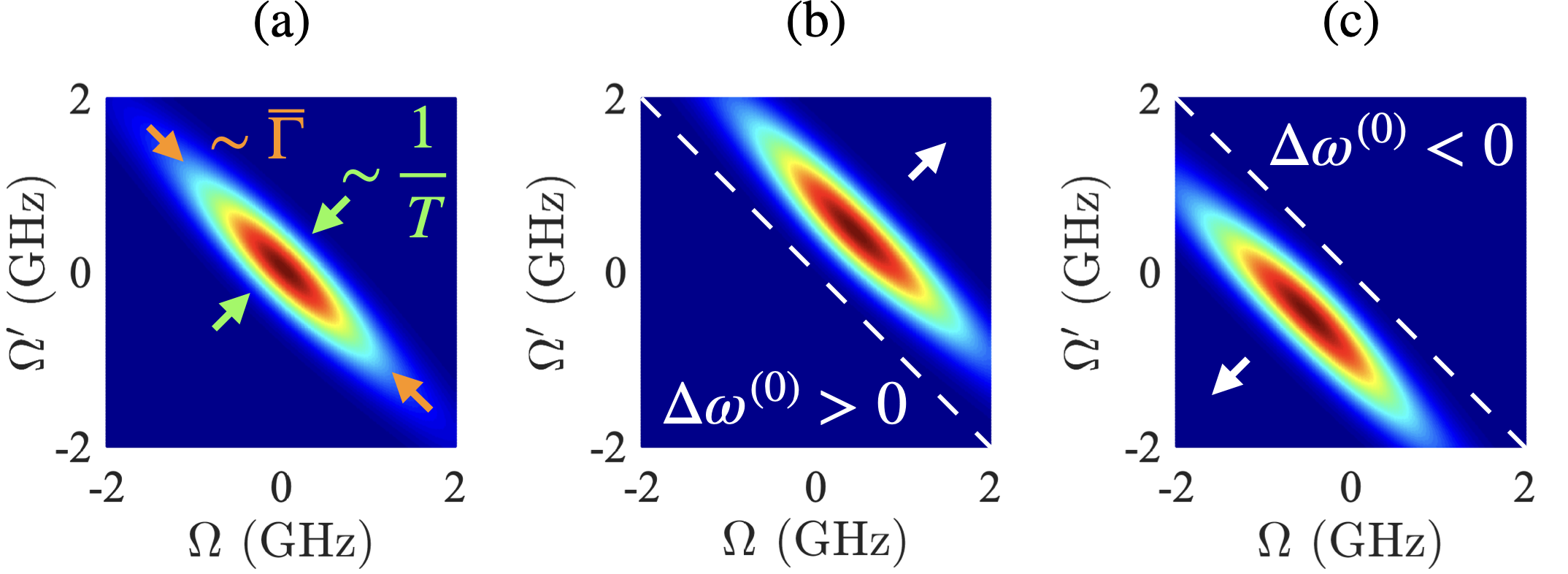}
    \caption{Demonstrating how a typical biphoton wavefunction amplitude for the first ring, $|\phi_1(\Omega, \Omega')|$, responds to changes in the system parameters, where (a) shows that the diagonal width (of size indicated by the green arrows) is inversely proportional to the effective pulse duration $\sim 1/T$ and the anti-diagonal width  (of size indicated by the orange arrows) is proportional to the resonance linewidth $\sim \overline{\Gamma}$.  (b) A frequency mismatch of $\Delta \omega^{(0)}_1 = 1\,{\rm GHz}$ causes a positive shift along the diagonal (in the direction of the white arrow), and (c) $\Delta \omega^{(0)}_1 =-1\,{\rm GHz}$  causes a negative  shift. The shifts $\pm 1 \,{\rm GHz}$ are exaggerated for illustrative purposes (see text). We assume the ring is critically coupled to the bus waveguide.}
    \label{fig:bpwf}
\end{figure}

Both pumps have Gaussian shapes centered on the pump frequencies $\omega^{(0)}_{P_a}$ and $\omega^{(0)}_{P_b}$, respectively. The pump amplitude incident on  ring $n$ (see Fig. \ref{fig:system}) is given by
\begin{align}
 \label{eq:pulse}
\alpha_n(\omega) = {\rm e}^{i\zeta_n}\sqrt{\frac{P_n}{\hbar \omega^{(0)}_P}}  T\exp(-\frac{1}{2}(\omega - \omega^{(0)}_P)^2T^2),
  \end{align}
where $\zeta_n$ is the pump phase, $P_n$ the incident peak power, $\omega^{(0)}_P$ the center frequency, and $T$ the effective pulse duration, taken to be the same for both pumps. Neglecting group velocity dispersion (GVD) near $\omega^{(0)}_P$, the pulse (\ref{eq:pulse}) corresponds to an incident power on each ring of $P_n(z,t) = P_n \exp(-(z/v - t)^2/T^2)$, where $z$ is the position in  the waveguide  and $v = v_{H}, v_{V}$ is the group velocity in the waveguide for $H$ or $V$ polarized light. We keep the peak power $P_n$ fixed as we vary $T$. 
As $T\rightarrow\infty$ the pulse (\ref{eq:pulse}) approaches the continuous-wave (cw) limit $\alpha_n(\omega) = \exp(i\zeta_n)\sqrt{2\pi P_n / (\hbar \omega^{(0)})}\, \delta(\omega - \omega_P^{(0)})$ as pointed out earlier \cite{MBanicRingLoss2022}. In the cw limit $T=\infty$ the constant incident cw power is given by $P_n(z,t) = P_n$. So the  peak power of the pulse in Eq. \eqref{eq:pulse}approaches the constant cw power in that limit.

In Appendix A
we calculate the  biphoton wavefunction describing the photon pair generated in each ring. A typical biphoton wavefunction amplitude $|\phi_1(\Omega, \Omega')|$ for ring 1 is shown in Fig. \ref{fig:bpwf} (a) for $T = 3\,{\rm ns}$, frequency $\omega^{(0)}_{P_a}$, and loaded quality factor $Q_{{\rm load},1} =  Q_{\rm int} / 2$ (i.e., critical coupling). In Fig. \ref{fig:bpwf} (a) the green and red arrows indicate that the width of $|\phi_1(\Omega,\Omega')|$ in the diagonal direction is inversely proportional to $T$, and the width in the anti-diagonal direction is proportional to the resonance linewidth $\overline{\Gamma} = \omega^{(0)}_{P_a}/(2 Q_{{\rm load},1})$.  For example, by overcoupling the ring to the waveguide (i.e., by increasing $\overline{\Gamma}$) the biphoton wavefunction is stretched along the anti-diagonal, and by decreasing $T$ it is stretched along the diagonal. Thus, by modifying the ring coupling and the pump pulse duration one can adjust each biphoton wavefunction and maximize the overlap integrals $O_{nn^\prime}$ in Eq. \eqref{eq:overlaps}.

The frequency mismatch between the resonances of ring 1 is defined as $\Delta\omega^{(0)}_1 \equiv 2\omega^{(0)}_{P_a} - \omega^{(0)}_{S_a} - \omega^{(0)}_{I_a}$, which can be approximated with $\Delta \omega^{(0)}_1 = -2 GVD_{P_a}(m_{P_a} - m_{S_a})^2/R_1^2$, where $GVD_{P_a} = d^2 \omega(k)/dk^2|_{K_{P_a}}$ is the GVD evaluated at the central wavenumber $ K_{P_a}$ for the pump, $R_1$ is the radius of ring 1, and $m_{P_a}$ and $m_{S_a}$ are the mode numbers for the pump and signal. For normal dispersion $GVD_{P_a} > 0 $ the frequency mismatch is negative $\Delta \omega^{(0)}_1<0$, while for anomalous dispersion $GVD_{P_a} < 0$ it is positive $\Delta \omega^{(0)}_1 >0$. For ring 1 we calculate $\Delta \omega^{(0)}_1 =  0.0166\,{\rm GHz}$, which will cause the biphoton wavefunction to slightly shift towards positive detuning frequency. In Fig. \ref{fig:bpwf} (b)  we  exaggerate this shift for illustrative purpose.  The frequency mismatch for each of the other rings is $\Delta \omega^{(0)}_2 = 0.0267\,{\rm GHz}$, $\Delta \omega^{(0)}_3 = 0.0174\,{\rm GHz}$, and $\Delta \omega^{(0)}_4 = 0.0279\,{\rm GHz}$, respectively. So each biphoton wavefunction is shifted towards positive frequency detuning by a slightly different amount, which can prevent their overlap integrals from equaling 1.

\begin{figure}
    \centering
    \includegraphics[width=\linewidth]{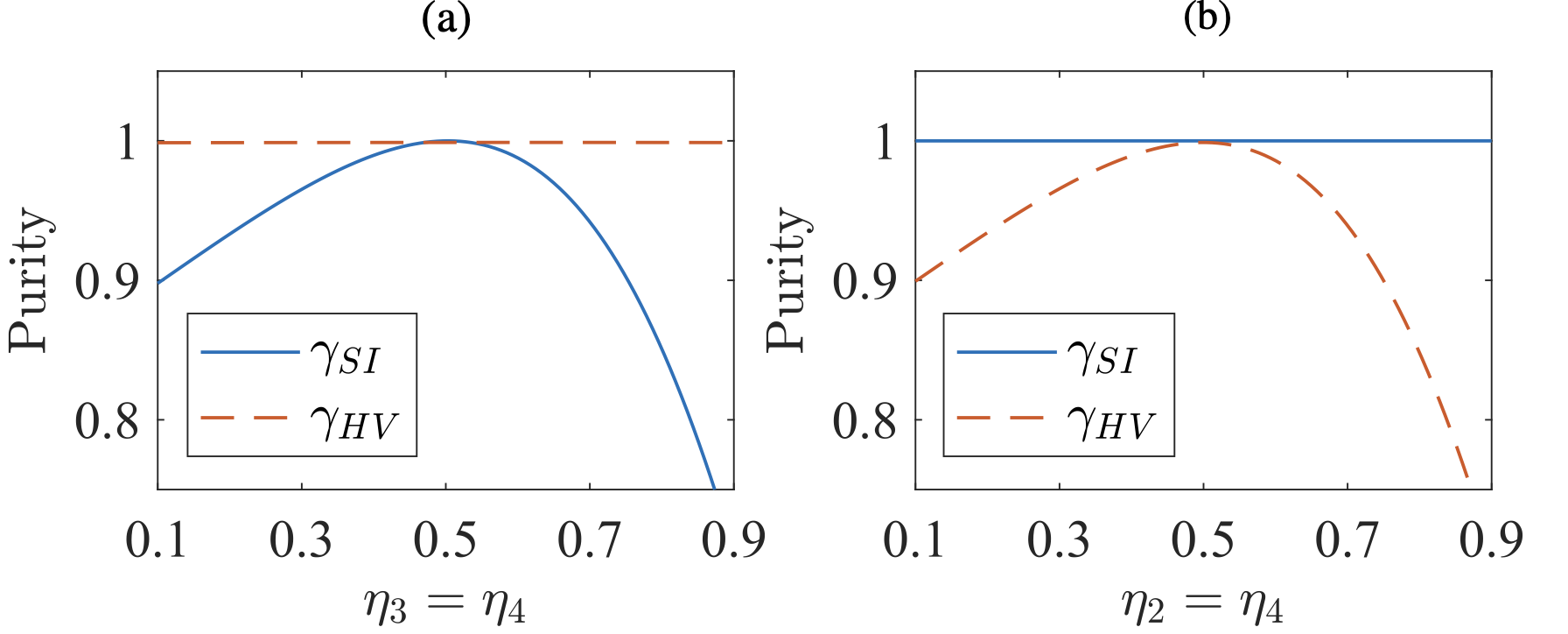}
    \caption{Purity of the reduced density operators for the frequency-bin DOF, $\gamma_{\rm bin}$ (blue line), and the polarization DOF, $\gamma_{\rm pol}$ (orange dashed), for (a) rings 1 and 2 critically coupled to the bus waveguide $\eta_1 =\eta_2 = 1/2$, as a function of equal coupling efficiencies of rings 3 and 4 to the bus waveguide ($\eta_3 = \eta_4$), and (b) rings 1 and 3 critically coupled to the bus waveguide $\eta_1 =\eta_3 = 1/2$, as a function of equal coupling efficiencies of rings 2 and 4 to the bus waveguide ($\eta_2 = \eta_4$). In (a) and (b)  we use  $T = 1\,{\rm ns}$ and assume an intrinsic quality factor of $Q_{\rm int} = 1\times10^6$.}
    \label{fig:purity vs eta}
\end{figure}

In Fig. \ref{fig:purity vs eta} (a) we show the purities $\gamma_{\rm bin}$ (blue line) and $\gamma_{\rm pol}$ (orange dashed) for rings 1 and 2 critically coupled,  and rings 3 and 4 having equal but variable coupling efficiencies. The coupling efficiency is defined as
\begin{align}
    \eta_n \equiv 1-\frac{Q_{{\rm load},n}}{Q_{\rm int}},
\end{align}
where $0\le\eta_n \le 1$. For \textit{critical coupling}  $\eta_n = 1/2$. In our calculations we put $\theta_3 = \theta_1 + \theta_2$,  $T = 1\,{\rm ns}$ for both pumps, and use peak powers around $1\,{\rm mW}$. In Fig. \ref{fig:purity vs eta} (a) the purity for either DOF is maximized when $\eta_3 = \eta_4 = 1/2$, which gives $\gamma_{\rm pol} = 99.942\%$ and $\gamma_{\rm bin} = 99.998\%$. Here $\gamma_{\rm pol}$ is insensitive to changes in $\eta_3 = \eta_4$, because it depends on the overlaps $O_{12}$ and $O_{34}$ involving rings with equal coupling efficiencies (i.e., $\eta_1 = \eta_2$ and $\eta_3 = \eta_4$). So we can expect $\phi_1(\Omega,\Omega')\approx \phi_2(\Omega,\Omega')$ and $\phi_3(\Omega,\Omega')\approx \phi_4(\Omega,\Omega')$. But, $\gamma_{\rm bin}$  is sensitive to changes in $\eta_3 = \eta_4$, because it depends on the overlaps $O_{13}$ and $O_{24}$ involving rings with  different coupling efficiencies. So we can expect $\phi_1(\Omega,\Omega')\neq \phi_3(\Omega,\Omega')$ and $\phi_2(\Omega,\Omega')\neq \phi_4(\Omega,\Omega')$. In Fig. \ref{fig:purity vs eta} (b) we consider a similar setup as above, except now $\eta_1 = \eta_3 = 1/2$ and we vary $\eta_2 = \eta_4$. Here we can expect $\phi_1(\Omega,\Omega')\approx \phi_3(\Omega,\Omega')$ and $\phi_2(\Omega,\Omega')\approx \phi_4(\Omega,\Omega')$, causing the purity $\gamma_{\rm bin}$ to be insensitive to changes in $\eta_2 = \eta_4$. As expected, to obtain the highest purity in both DOFs the coupling efficiencies of all rings should be similar.

\begin{figure}
    \centering
    \includegraphics[width=0.8\linewidth]{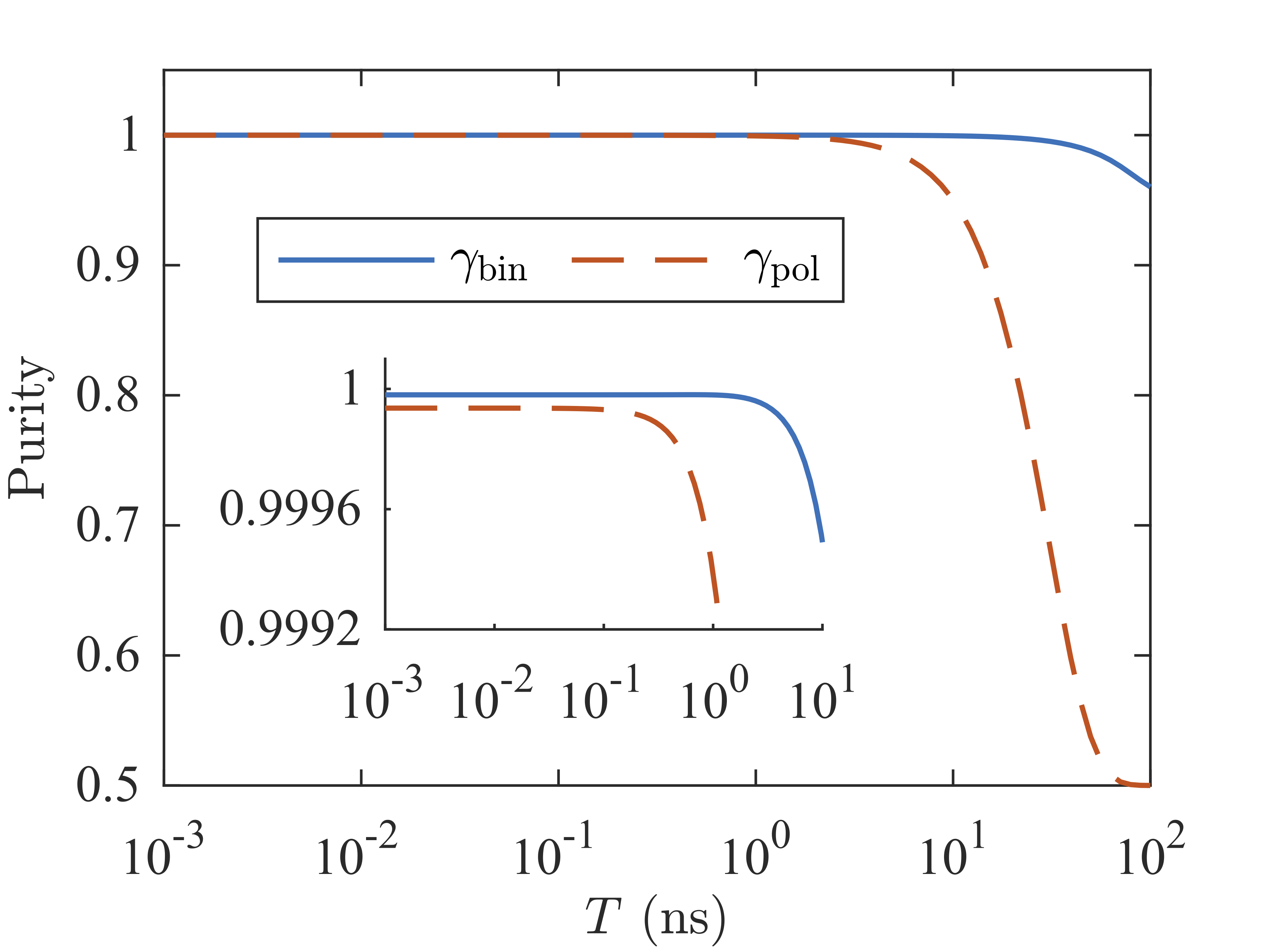}
    \caption{Purity of the reduced density operators for the frequency-bin DOF, $\gamma_{\rm bin}$ (blue line), and the polarization DOF, $\gamma_{\rm pol}$ (orange dashed), for all rings critcally coupled to the bus waveguide $\eta_1 =\eta_2 =\eta_3 = \eta_4 = 1/2$, as a function of $T$. We assume an intrinsic quality factor of $Q_{\rm int} = 1\times 10^6$.}
    \label{fig:purity vs tau}
\end{figure}

In Fig. \ref{fig:purity vs tau}, we show $\gamma_{\rm pol}$ (orange dashed) and $\gamma_{\rm bin}$ (blue line) for increasing $T$, by considering all rings in critical coupling. For $T$ much shorter than the dwelling time of the photon ($\sim 5\,{\rm ns}$) the width of each biphoton wavefunction ($\sim 1/T$) is generally much larger than the shift of its center caused by $\Delta \omega^{(0)}>0$. The relative shift between two biphoton wavefunctions for a longer $T$, however, can be on the order of their widths. So generally  as $T$ increases the overlap integrals (\ref{eq:overlaps}) approach zero and the purity decreases. For $T=1\,{\rm ns}$ we obtain $\gamma_{\rm pol} = 99.939\%$ and $\gamma_{\rm bin} = 99.998\%$ (see inset of Fig. \ref{fig:purity vs tau}), and for $T=10\,{\rm ns}$,  $\gamma_{\rm pol} = 95.021\%$ (a $5\%$ decrease) and $\gamma_{\rm bin} = 99.949\%$ (a $0.05\%$ decrease). Increasing $T$ causes $\gamma_{\rm pol}$ to decrease more than $\gamma_{\rm bin}$, because the relative shift in the centers of the biphoton wavefunctions from rings 1 and 2 ($|\Delta \omega^{(0)}_1 -\Delta \omega^{(0)}_2| \approx 0.01\,{\rm GHz}$) is an order of magnitude larger than it is for the biphoton wavefunctions from rings 1 and 3 ($|\Delta \omega^{(0)}_1 -\Delta \omega^{(0)}_3| \approx 0.001\,{\rm GHz}$).

To achieve a high degree of purity in both DOFs in our system the coupling efficiencies of all the rings should be the same and $T$ should not exceed $1\,{\rm ns}$. However, we show in the next section that the generation rate of photon-pairs is large for long pulses. So there is a trade-off between high purity in both DOFs and a large generation rate. 

\section{Generation rate of hyperentangled photon pairs}

\begin{figure}
    \centering
    \includegraphics[width=\linewidth]{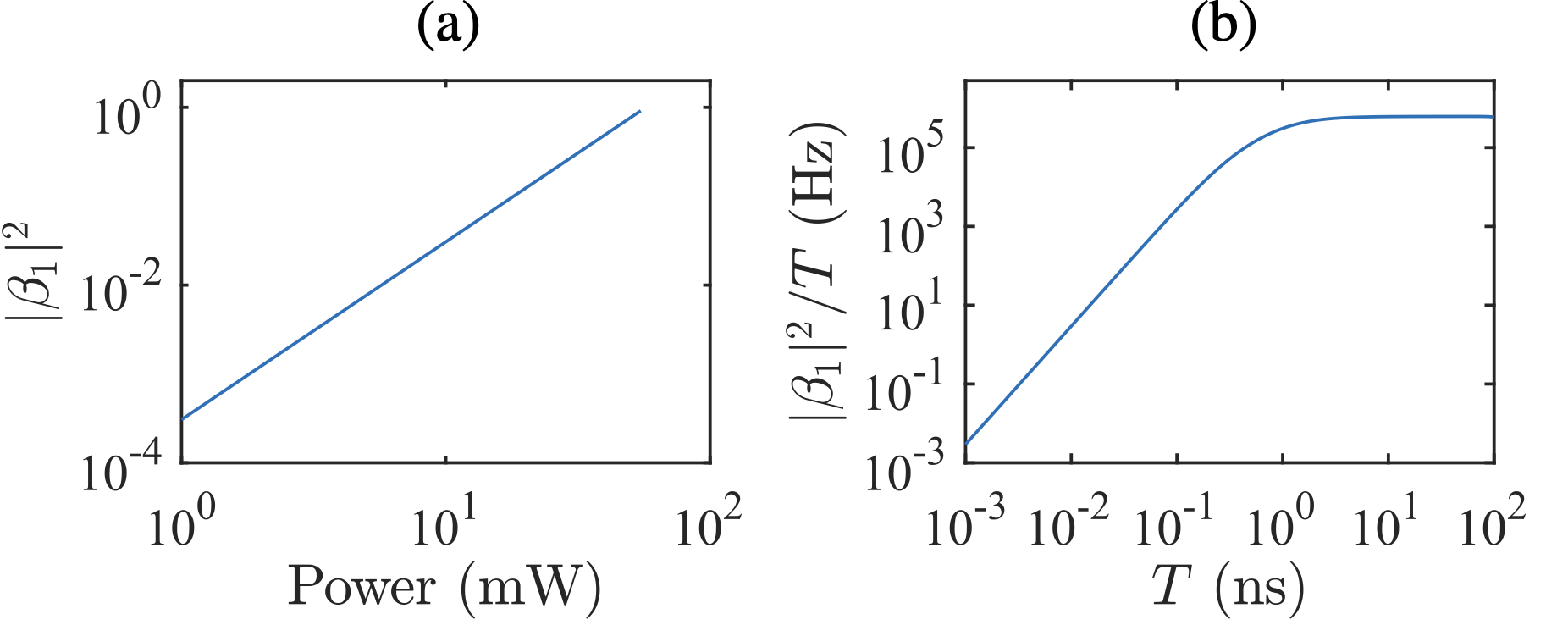}
    \caption{Characterizing the photon-pair production in ring 1, with (a) the pair-generation probability $|\beta_1|^2$ as a function of incident peak pump power, and (b) the pair-generation rate $|\beta_1|^2/T$ versus $T$ for a peak power of $1\,{\rm mW}$. Both (a) and (b) are with critical coupling $\eta_1 = 1/2$ and assuming an intrinsic quality factor of $Q_{\rm int} = 1\times 10^6$.}
    \label{fig:beta plot}
\end{figure}
In this section, we analyze the efficiency of the system. Here, we validate this assumption by examining the generation probability for each ring individually, ensuring that the photon pair generation remains within acceptable and practical limits.

To start, we investigate the validity of our assumption that each ring generates at most a pair of photons. First, we consider the state generated by a single ring $n$ and impose the condition that the probability of generating a pair from this ring is small, $|\beta_n|^2 \ll 1$. In Fig. \ref{fig:beta plot} (a), we present $|\beta_1|^2$ for ring 1 as a function of the peak power incident on the ring, assuming $T=1,{\rm ns}$ and critical coupling, calculated as described in Appendix A. In the results section (Sec. \ref{sec:results HE}), where all rings are critically coupled, we set the peak pump power incident on each ring to approximately $1,{\rm mW}$. As shown in Fig. \ref{fig:beta plot} (a), a peak pump power of $1,{\rm mW}$ corresponds to $|\beta_1|^2 = 3\times10^{-4}$, satisfying the condition $|\beta_1|^2 \ll 1$. This result confirms that the assumption of low photon pair generation probability holds for each ring, supporting the overall model used in our analysis.

The rate at which photon pairs are generated by a given ring is the probability of generating a pair divided by the effective pulse duration, $\mathcal{R}_n = |\beta_n|^2/T$. 
In Fig. \ref{fig:beta plot} (b) we plot the pair-generation rate for ring 1, as a function of $T$, for an incident peak power of $1\,{\rm mW}$. In that figure the rate for $T=1\,{\rm ns}$ is $\mathcal{R}_1 = 3.07\times10^5\,{\rm Hz}$, and given that we have equalized the rate from each ring, the total generation rate for the system is $\mathcal{R}_{\rm tot} \equiv 4 \mathcal{R}_1 = 1.23\times 10^6\,{\rm Hz}$. Using $T=100\,{\rm ns}$ instead we roughly double the rate ($5.97\times10^5\,{\rm Hz}$) but the purity decreases by approximately $4\%$ for the bin DOF and $50\%$ for the polarization DOF.  

\section{Conclusion}
We have proposed an integrated circuit design for generating polarization-frequency-bin hyperentangled photon pairs. Both the polarization and frequency-bin degrees of freedom (DOF) of the photon pairs are entangled on-chip through spontaneous four-wave mixing (SFWM) interactions in a series of ring resonators. A programmable sequence of Mach-Zehnder interferometers (MZIs) and add-drop filters is used to control the intensity and phase of the pump pulse incident on each ring. We derived three conditions that must be satisfied to achieve hyperentanglement. First, the pair-generation probability must be equal across all rings. Second, the relative phase between the pumps incident on different rings must maintain a fixed relationship. Third, the biphoton wavefunctions for each ring should be identical, meaning that the continuous frequency DOF characteristics of the photon pairs must be the same for each ring. We demonstrated numerically that these three conditions can be met under realistic system parameters, accounting for variations in the resonance linewidths and dispersion of each ring, and we obtained a generation rate for the hyperentangled state of $10^6,{\rm Hz}$. Thus, this approach can efficiently generate on-chip polarization-frequency-bin hyperentanglement, and we believe it will be of great interest to those developing quantum information processing with photonic integrated circuits.

\section*{Acknowledgements}
The authors would like to acknowledge the Horizon-Europe research and innovation program under Grant Agreement No. 101070168
(HYPERSPACE) for funding this work. J. E. S. and C. V. acknowledge Natural  Sciences  and  Engineering  Research Council of Canada for funding.

\bibliography{References-minimum}

\appendix

\section{Appendix A: The Biphoton Wavefunction}
\label{sec:biphoton_wavefunction}

For the specific case of photon pairs generated by ring resonators, the biphoton wavefunction in a given ring $n$ assuming that the group velocities at the pump, signal, and idler frequencies are approximately equal can be written as \cite{MBanicRingLoss2022} 
\begin{align}
\label{eq:phi1}
\phi_n(\omega,\omega')&=i\hbar\frac{\alpha^2_{n}}{ \beta_n} \Lambda_n R_n \sqrt{\omega^{(0)}_{S}\omega^{(0)}_{I}}F^*_{nS +}(\omega) F^*_{nI +}(\omega')\nonumber
\\
&\times g_n(\omega,\omega'),
\end{align}
where $\Lambda_n$ is the nonlinear parameter for SFWM in the ring $n$, $R_n$ is the ring radius, and $\omega^{(0)}_{S}$ and $\omega^{(0)}_{I}$ are the center frequencies of the signal and idler resonance in the ring, where $S = S_a, S_b$ and $I = I_a,I_b$. Here $\alpha_n$ is the pump amplitude incident on ring $n$, given by
\begin{align}
\label{eq:alpha n}
\alpha_n \equiv {\rm e}^{i\zeta_n} \sqrt{\frac{P_n}{\hbar \omega^{(0)}_P}},
\end{align}
where $\zeta_n$ is phase of the pump just before it enters the ring, $P_n$ is the peak power of the pump incident to the ring, and $\omega^{(0)}_P$ is the center frequency of the resonance for the pump, where $P = P_a, P_b$.  In Eq. \eqref{eq:phi1} we have introduced the function $g_n(\omega, \omega')$ given by
\begin{align}
\label{eq:g}
g_n(\omega,\omega') &= \int d\omega_1 F_{nP-}(\omega_1) F_{nP-}(-\omega_1+\omega +\omega') \nonumber
\\
&\times A(\omega_1) A(-\omega_1+\omega +\omega'),
\end{align}
where the field enhancement factors for the ring $F_{nJ}(\omega)$, where $J=P_a,P_b,S_a,S_b,I_a,I_b$, can be written as
\begin{align}
\label{eq:F}
F_{nJ\pm}(\omega) &= -i \sqrt{\frac{v_n \eta_n }{\pi R_n\overline{\Gamma}_n}} \frac{1}{i(\omega - \omega^{(0)}_J)/\overline{\Gamma}_n \pm 1},
 \end{align}
where $v_n$ is the group velocity, $\overline{\Gamma}_n$ is the resonance linewidth, $\omega^{(0)}_J$ is the center frequency of the resonance, and $\eta_n$ is the coupling efficiency of the light in the ring to waveguide I (see Fig. \ref{fig:system}). Also we have introduced the  pump pulse function $A(\omega)$ as
 \begin{align}
 \label{eq:pulse A}
A(\omega) =   T\exp(-\frac{1}{2}(\omega - \omega^{(0)}_P)^2T^2),
  \end{align}
 where we define $T$ as the effective pulse duration in time. As $T\rightarrow \infty$ we obtain the cw limit $A(\omega) \rightarrow \sqrt{2\pi}\delta(\omega - \omega^{(0)}_P)$ as pointed out earlier by Banic \textit{et al.} \cite{MBanicRingLoss2022}. Eq. \eqref{eq:pulse} for the pump pulse amplitude $\alpha_n(\omega)$ is given by  
 \begin{align}
     \alpha_n(\omega) = \alpha_n A(\omega),
 \end{align}
 where $\alpha_n$ is given by Eq. \eqref{eq:alpha n}.
 
 The magnitude $|\beta_n|$ is obtained from the requirement that the biphoton wavefunction is normalized.   We find
 \begin{align}
 \label{eq:norm1}
     |\beta_n|^2 &= \hbar^2|\alpha_n|^4 |\Lambda_n|^2 R^2_n \omega^{(0)}_{S}\omega^{(0)}_{I}\nonumber
     \\
     &\times \int d\omega d\omega' |F_{nS +}(\omega)|^2 |F_{nI +}(\omega')|^2 |g_n(\omega,\omega')|^2,
 \end{align}
which gives the probability that a photon pair is generated by ring $n$. 

 Introducing the detuning frequency $\Omega \equiv \omega - \omega^{(0)}_J$, we can write Eq. \eqref{eq:phi1} for the biphoton wavefunction as
\begin{align}
\label{eq:phi1 detuned}
\phi_n(\Omega,\Omega')&=i\hbar\frac{\alpha^2_{n}}{ \beta_n} \Lambda_n R_n \sqrt{\omega^{(0)}_{S}\omega^{(0)}_{I}}F^*_{nS +}(\Omega) F^*_{nI +}(\Omega') \nonumber
\\
&\times g_n(\Omega,\Omega'),
\end{align}
where 
\begin{align}
\label{eq:g detuned}
g_n(\Omega,\Omega') &= \int d\Omega_1  F_{nP-}(-\Omega_1+\Omega +\Omega' - \Delta \omega^{(0)}) \nonumber
\\
&\times F_{nP-}(\Omega_1)A(\Omega_1) A(-\Omega_1+\Omega +\Omega' - \Delta \omega^{(0)}),
\end{align}
where we have introduced the frequency mismatch between the center frequencies of the pump, signal, and idler resonances as $\Delta \omega^{(0)} = 2\omega^{(0)}_P - \omega^{(0)}_S - \omega^{(0)}_I$. The field enhancement factors are now given by
\begin{align}
\label{eq:F detuned}
F_{nJ\pm}(\Omega) &= -i \sqrt{\frac{v_n \eta_n }{\pi R_n\overline{\Gamma}_n}} \frac{1}{i\Omega/\overline{\Gamma}_n \pm 1},
 \end{align}
and finally the pump pulse function is given by
 \begin{align}
 \label{eq:pulse detuned}
 A(\Omega) =  T\exp(-\frac{1}{2}\Omega ^2T^2).
  \end{align}


\section{Appendix B: Separating DOFs}
\label{sec:separation}

In this appendix we show that the composite ket in  Eq. \eqref{eq:composite ket}  can be written as a product of three single-DOF kets: one for the polarization DOF, one for the frequency-bin DOF, and one for the continuous-frequency detuning DOF. We show that Eq. \eqref{eq:coefficients} for the coefficients $c_{BPB'P'}(\Omega, \Omega')$ are given by Eq. \eqref{eq:coefficients solution}.

The eigenvalue equation for $\Sigma_{0B}$ is $\Sigma_{0B} \ket{P,B';\Omega} = \delta_{BB'}\ket{P,B';\Omega}$, where the eigenkets $\ket{P,B;\Omega}$ are given by Eq. \eqref{eq:composite ket}. Putting Eq. \eqref{eq:eigenket expand} for the expansion of $\ket{P,B;\Omega}$ into both sides of this eigenvalue equation and using Eq. \eqref{eq:bin decompose} for the decomposition of $\Sigma_{0B}$ we find that we can write $c_{BPB'P'}(\Omega, \Omega')$  as
\begin{align}
\label{eq:cbin}
    c_{BPB'P'}(\Omega, \Omega') = \delta_{BB'}\overline{c}_{PP'B}(\Omega, \Omega'),
\end{align}
where $\overline{c}_{PP'B}(\Omega, \Omega')$ is a function we will determine.

The eigenvalue equations for $\hat{P}$ are $\hat{P} \ket{H,B;\Omega} = \ket{H,B;\Omega}$ and $\hat{P} \ket{V,B;\Omega} = -\ket{V,B;\Omega}$. Putting Eq. \eqref{eq:eigenket expand} for the expansion of $\ket{P,B;\Omega}$ into both sides of these eigenvalue equations and using Eq. \eqref{eq:pol decompose} for the decomposition of $\hat{P}$ we find that we can write $\overline{c}_{PP'B}(\Omega, \Omega')$  as
\begin{align}
\label{eq:cpol}
   \overline{c}_{PP'B}(\Omega, \Omega') = \delta_{PP'}\widetilde{c}_{PB}(\Omega, \Omega'),
\end{align}
where $\widetilde{c}_{PB}(\Omega, \Omega')$ is a function we will determine.

The eigenvalue equation for $\hat{\Omega}$ is $\hat{\Omega} \ket{P,B;\Omega} = \Omega \ket{P,B;\Omega}$. Putting Eq. \eqref{eq:eigenket expand} for the expansion of $\ket{P,B;\Omega}$ into both sides of this eigenvalue equation and using Eq. \eqref{eq:freq decompose} for the decomposition of $\hat{\Omega}$ we find that we can write $\widetilde{c}_{PB}(\Omega, \Omega')$  as
\begin{align}
\label{eq:cfreq}
   \widetilde{c}_{PB}(\Omega, \Omega') = \delta(\Omega - \Omega')f_{PB}(\Omega),
\end{align}
where $f_{PB}(\Omega)$ is a function we will determine.

Combining Eq. \eqref{eq:cbin}, Eq. \eqref{eq:cpol}, and Eq. \eqref{eq:cfreq} we obtain
\begin{align}
\label{eq:c final}
     c_{BPB'P'}(\Omega, \Omega') = f_{PB}(\Omega)\delta_{PP'}\delta_{BB'}\delta(\Omega - \Omega').
\end{align}
Putting Eq. \eqref{eq:c final} into Eq. \eqref{eq:eigenket expand} for the expansion of the eigenkets
\begin{align}
    \ket{P, B; \Omega} = f_{PB}(\Omega)\ket{P} \otimes \ket{B} \otimes \ket{\Omega}.
\end{align}
Now to preserve the inner product between two composite kets we have that
\begin{align}
    |f_{PB}(\Omega)|^2 = 1.
\end{align}
Since $f_{PB}(\Omega)$ is just a phase that we can absorb into the biphoton wavefunction we neglect it and put $f_{PB}(\Omega) = 1$.

%

\section{Appendix C: Reduced density operator for polarization and frequency-bin DOFs}
\label{sec:reduced density operator}
In this appendix we calculate the reduced density operator $\overline{\rho}$ that results from tracing over the continuous-frequency detuning variables $\Omega$ and $\Omega'$. The reduced density operator is calculated with
\begin{align}
\label{eq:red_rho}
    \overline{\rho} = \int d\Omega d\Omega' \bra{\Omega\Omega'}\ket{\psi}\bra{\psi}\ket{\Omega\Omega'},
\end{align}
where $\ket{\psi}$ is given by Eq. \eqref{eq:4 ring state 2} for the state generated by the system.
Putting Eq. \eqref{eq:4 ring state 2} into  Eq. \eqref{eq:red_rho} we obtain
\begin{align}
\label{eq:red_rho 16 terms}
    \overline{\rho} & = \frac{1}{4}( \ket{aa}\bra{aa} + U^*_2 O_{13}\ket{aa}\bra{bb}+ U_2 O^*_{13}\ket{bb}\bra{aa}\nonumber
    \\
    &+ \ket{bb}\bra{bb})\otimes\ket{HH}\bra{HH} \nonumber
    \\
    &+\frac{1}{4}( \ket{aa}\bra{aa} +U_1 U^*_3 O_{24}\ket{aa}\bra{bb}+ U_3U^*_1 O^*_{24}\ket{bb}\bra{aa}\nonumber
    \\
    &+ \ket{bb}\bra{bb})\otimes\ket{VV}\bra{VV} \nonumber
    \\
    &+\frac{1}{4}( U^*_1O_{12}\ket{aa}\bra{aa} + U^*_3 O_{14}\ket{aa}\bra{bb} \nonumber
    \\
    &+ U^*_1U_2 O_{32}\ket{bb}\bra{aa}+ U^*_3U_2O_{34}\ket{bb}\bra{bb})\otimes\ket{HH}\bra{VV} \nonumber
    \\
    &+\frac{1}{4}( U_1O^*_{12}\ket{aa}\bra{aa} + U_3 O^*_{14}\ket{bb}\bra{aa} \nonumber
    \\
    &+ U_1U^*_2 O^*_{32}\ket{aa}\bra{bb}+ U_3U^*_2O^*_{34}\ket{bb}\bra{bb})\otimes \ket{VV}\bra{HH},
\end{align}
where we used Eq. \eqref{eq:inner prod freq} for the inner product between two kets $\ket{\Omega}$ and $\ket{\Omega'}$ and Eq. \eqref{eq:overlaps} for the overlap integrals $O_{nn'}$ between the biphoton wavefunctions of rings $n$ and $n'$. For convenience we have  defined the unit complex coefficients
\begin{align}
    U_1 \equiv {\rm e}^{2i\theta_1},\,\,\,\,U_2\equiv {\rm e}^{2i \theta_2},\,\,\,U_3\equiv {\rm e}^{2i\theta_3},
\end{align}
and the simplified frequency-bin kets
\begin{align}
    \ket{a a} \equiv \ket{S_a I_a},\,\,\,\ket{b b} \equiv \ket{S_b I_b}.
\end{align}

\end{document}